\documentclass[a4paper,12pt]{article}

\pdfoutput=1

\usepackage{amsmath}
\usepackage{amssymb}
\usepackage{graphicx}
\usepackage{hyperref}
\usepackage{cite}
\usepackage{color}

\makeatletter
\@addtoreset{equation}{section}
\renewcommand{\theequation}{\thesection.\@arabic\c@equation}
\makeatother

\begin{document}

\begin{titlepage}

\vspace*{-15mm}   
\baselineskip 10pt   
\begin{flushright}   
\begin{tabular}{r}    
{\tt NU-QG-6}\\   
\end{tabular}   
\end{flushright}   
\baselineskip 24pt   
\vglue 10mm   

\begin{center}
{\Large\bf
 Self-gravitating strings and quantum effects in two-dimensional gravity
}

\vspace{8mm}   

\baselineskip 18pt   

\renewcommand{\thefootnote}{\fnsymbol{footnote}}

Akihiro~Ishibashi${}^{1,2,3,4}$\footnote[2]{ishibashi.akihiro.r7@f.mail.nagoya-u.ac.jp}, 
Yoshinori~Matsuo${}^{1,3}$\footnote[3]{ymatsuo@nagoya-u.jp} 
and 
Akane~Tanaka${}^{3}$\footnote[4]{atanaka.bh@gmail.com} 

\vspace{5mm}   

{\it  
$^1$
Department of Physics, Nagoya University, Nagoya 464-6802, Japan
\\
$^2$
Kobayashi-Maskawa Institute, Nagoya University, Nagoya 464-8602, Japan
\\
$^3$ Department of Physics and $^4$Research Institute for Science and Technology, \\ 
Kindai University, Higashi-Osaka, Osaka 577-8502, Japan
}

\renewcommand{\thefootnote}{\arabic{footnote}}
  
\vspace{10mm}   

\end{center}

\begin{abstract}
It is expected that when the string coupling is taken to be sufficiently small, a black hole turns into a bound state of self-gravitating fundamental strings. This state would be described by winding strings wrapping around the Euclidean time circle, known as the Horowitz-Polchinski solution. In this paper, we study such a self-gravitating string configuration in two-dimensional dilaton gravity theories. 
We first derive an analytic expression of the solution describing a winding string in two-dimensions and investigate in detail the geometry of this solution. Our winding string solution in two dimensions describes the geometry near the surface of the bound state of self-gravitating strings in the large-dimension limit, much like a two-dimensional black hole describes the near horizon geometry of the Schwarzschild black hole in the large-dimension limit. To study quantum effects around self-gravitating strings, we obtain an analytic solution of winding strings in the RST model. In a similar fashion to the Hartle-Hawking vacuum around a black hole, our string solution in the RST model contains the background radiation, whose temperature 
is the same as that of the winding strings. 
\end{abstract}

\baselineskip 18pt   

\end{titlepage}

\newpage

\baselineskip 18pt

\noindent\rule{\textwidth}{1pt}

\tableofcontents

\noindent\rule{\textwidth}{1pt}


\section{Introduction}\label{sec:Intro}

The black hole information loss paradox is a fundamental problem in quantum gravity. 
At the final stages of the black hole evaporation, 
the size and temperature of the black hole become comparable to the Planck scale, 
and quantum effects of gravity will play an important role. 
String theory is one of the promising candidates of a quantum theory of gravity, 
and is expected to give a microscopic description of black holes. 

Susskind proposed that black holes transit into strings when the size becomes comparable to the string scale \cite{Susskind:1993ws}%
\footnote{%
See \cite{Susskind:2021nqs} for a brief review and 
\cite{Bowick:1985af,Horowitz:1996nw,Horowitz:1997jc,Damour:1999aw,Khuri:1999ez,Barbon:2004dd,Giveon:2006pr,Mertens:2014dia,Kawamoto:2015zha,Brustein:2021ifl,Chen:2021emg,Chen:2021dsw,Brustein:2021lnr,Brustein:2021qkj,Brustein:2022uft,Balthazar:2022szl,Brustein:2022wiq,Balthazar:2022hno,Ceplak:2023afb,Santos:2024ycg} for related works.} 
Such small black holes have the Hawking temperature near the Hagedorn temperature. 
In order to study black holes at high temperatures, winding modes of strings play an important role. 
Beyond the Hagedorn temperature, more and more massive strings are excited by the thermal energy. 
These massive strings can be described by winding strings wrapping on the Euclidean time circle. 
Horowitz and Polchinski formulated a model of self-gravitating strings by using the winding strings, 
which describes the string phase of black holes after the transition \cite{Horowitz:1997jc}. 
The Euclidean geometry in the string phase takes the form of an annulus 
while the black hole spacetime is given by the cigar geometry as 
the Euclidean time circle shrinks to a point at the horizon. 

In this paper, we study the black hole/string transition 
and quantum effects in the two-dimensional dilaton gravity 
\cite{Witten:1991yr,Elitzur:1990ubs,Mandal:1991tz,Dijkgraaf:1991ba,Callan:1992rs}. 
The cigar geometry of the black hole appears as the target space of 
the gauged WZW model of the $SL(2,\mathbb R)_k/U(1)$ coset \cite{Witten:1991yr}. 
It can also be obtained by taking the large-dimension limit of the Schwarzschild spacetime 
by focusing on an infinitesimal region near the horizon \cite{Soda:1993xc,Emparan:2013xia}. 
The black hole in two-dimensional gravity 
has also been studied as a toy model of the black hole evaporation \cite{Callan:1992rs}. 
A small modification is introduced in \cite{Russo:1992ax} 
to solve the model exactly incorporating quantum effects, which is known as the RST model. 
Since the two-dimensional black hole above appears 
as a target space of the non-critical string theory, 
winding strings can also be introduced straightforwardly. 
The classical solution of winding strings, 
which corresponds to the Horowitz-Polchinski solution, 
in the two-dimensional gravity was studied numerically in \cite{Brustein:2021qkj}. 
In this paper, we first derive an analytic expression of the solution 
to see more details of the geometry in the presence of the winding strings. 
Then, we calculate the Horowitz-Polchinski solution in the RST model 
to see the quantum correction around the self-gravitating strings. 

This paper is organized as follows. 
In Sec.~\ref{sec:cls}, we study self-gravitating strings 
in the classical two-dimensional dilaton gravity. 
We briefly review the two-dimensional dilaton gravity in Sec.~\ref{ssec:cls-BH} 
and the effective theory of winding strings in Sec.~\ref{ssec:cls-str}. 
In \cite{Brustein:2021qkj}, this effective theory of winding strings is solved 
only numerically and analytic expression is given only for asymptotic behavior near the inner boundary. 
In this paper, we give an analytic expression of the entire solution 
and discuss the structure of the spacetime in the presence of winding strings in Sec.~\ref{ssec:cls-sol}. 
In Sec.~\ref{sec:RST}, we study self-gravitating strings in the RST model. 
We first review the RST model in Sec.~\ref{ssec:RST-BH}, 
and then, introduce winding strings in Sec.~\ref{ssec:RST-str}. 
As in the RST model, we also make small modifications in the action of the winding string field, 
which vanish in the classical limit. 
In Sec.~\ref{ssec:RST-sol}, we derive the winding string solution in the RST model. 
In Sec.~\ref{ssec:Entropy}, we study the entropy and free energy of winding strings in the RST model. 
Sec.~\ref{sec:Conclusion} is devoted to the conclusion and discussion.


\section{Self-gravitating strings in two dimensions}
\label{sec:cls}


\subsection{Black hole in two-dimensional dilaton gravity}
\label{ssec:cls-BH}

We first review the black hole solution in the two-dimensional dilaton gravity 
\cite{Witten:1991yr,Elitzur:1990ubs,Mandal:1991tz,Dijkgraaf:1991ba,Callan:1992rs}. 
The action is given by 
\begin{align} 
 \mathcal I 
 &= 
 \frac{1}{2\pi} \int d^2x \sqrt{-g}\,e^{-2\phi}
 \left[
  R + 4(\partial\phi)^2 + \frac{4}{k}
 \right] 
 + \mathcal I_\text{matter} 
\ , 
\label{CGHS} 
\end{align}
where $R$ is the scalar curvature and $\phi$ is the dilaton. 
In the conformal gauge, 
\begin{equation}
 ds^2 = - e^{2\rho} dx_+ dx_- \ , 
\label{metric}
\end{equation}
the equations of motion for the metric are expressed as 
\begin{align}
 0 
 &= 
 e^{-2\phi} 
 \left[
  4 \left(\partial_+ \rho\right) \left(\partial_+ \phi\right) - 2 \partial_+^2 \phi
 \right] 
 + T_{++} \ , 
\\
 0 
 &= 
 e^{-2\phi} 
 \left[
  4 \left(\partial_- \rho\right) \left(\partial_- \phi\right)  - 2 \partial_-^2 \phi
 \right] 
 + T_{--} \ , 
\\
 0 
 &= 
 e^{-2\phi} 
 \left[
  2 \partial_+ \partial_- \phi 
  - 4 \left(\partial_+\phi\right) \left(\partial_-\phi\right)
  - \frac{1}{k} e^{2\rho} 
 \right] 
 + T_{+-} \ , 
\end{align}
where $T_{\mu\nu}$ is the energy-momentum tensor of the matter, defined by%
\footnote{%
Note that this convention in two-dimensional gravity is 
different from the standard definition by a factor $\pi$. 
} 
\begin{equation}
 T_{\mu\nu} = - \frac{2\pi}{\sqrt{-g}} \frac{\delta\mathcal I_\text{matter}}{\delta g^{\mu\nu}} \ , 
\label{EMtensor}
\end{equation}
and for the dilaton, 
\begin{equation}
 0 = 
 - 4 \partial_+ \partial_- \phi + 4 \left(\partial_+\phi\right) \left(\partial_-\phi\right) 
 + 2 \partial_+ \partial_- \rho + \frac{1}{k} e^{2\rho} + e^{2\rho+2\phi} L \ , 
\end{equation}
where 
\begin{equation}
 L = - \frac{\pi}{4\sqrt{-g}} \frac{\delta \mathcal I_\text{matter}}{\delta\phi} \ . 
\label{L}
\end{equation}

In the absence of the matter, these equations reduce to 
\begin{align}
 0 &= \partial_+ \partial_- e^{-2\phi} + \frac{1}{k} e^{2\rho-2\phi} \ , 
\\
 0 &= \partial_+ \partial_- \left(\rho - \phi\right) \ . 
\end{align}
The solution of the second equation is given by 
a sum of the holomorphic and anti-holomorphic functions as 
\begin{equation}
 \rho - \phi = F_+(x_+) + F_-(x_-) \ . 
\end{equation}
Since $\phi$ is invariant under coordinate transformations and 
$\rho$ transforms as 
\begin{equation} 
\label{transf:rho}
 \rho \to \rho - \frac{1}{2} \log f_+(x_+) - \frac{1}{2} \log f_-(x_-) \ , 
\end{equation}
where 
\begin{equation}
 f_\pm(x_\pm) = \frac{dw_\pm}{dx_\pm} \ , 
\end{equation}
for the conformal coordinate transformation $x_\pm \to w_\pm(x_\pm)$, 
we can always take $\rho = \phi$ by choosing appropriate coordinates. 
Then, the equations of motion become 
\begin{align}
 0 &= 
 \partial_\pm^2 e^{-2\phi} \ , 
\\
 0 &= 
 \partial_+ \partial_- e^{-2\phi} + \frac{1}{k} \ . 
\end{align}
The solution of these equations is given by 
\begin{equation}
 e^{-2\phi} = M - \frac{1}{k} x_+ x_- \ , 
\label{CGHSBH}
\end{equation}
up to constant shifts of coordinates $x_+$ and $x_-$, where $M$ is a constant. 
Thus, the metric is expressed as 
\begin{equation}
 ds^2 = - \frac{k\ dx_+dx_-}{1-x_+x_-} \ , 
\label{2dBH}
\end{equation}
up to an appropriate rescaling of $x_+$ and $x_-$. 
This is nothing but a black hole geometry, whose 
singularity and horizon are located at $x_+ x_- = 1$ and $x_+ x_- = 0$, respectively. 
The spacetime is static and asymptotically flat. 
In order to see this, it is convenient to introduce 
the time and tortoise coordinates $(t,r_*)$ as 
\begin{equation}
 x_\pm = \pm\exp\left[\frac{r_* \pm t}{\sqrt k}\right] \ , 
\end{equation}
and the metric is expressed as 
\begin{equation}
 ds^2 = - \left(1+e^{-2r_*/\sqrt k}\right)^{-1} \left(-dt^2 + dr_*^2\right) \ . 
\label{2dBHrs}
\end{equation}
We further define the radial coordinate $x$ by 
\begin{equation}
 \left(1+e^{-2r_*/\sqrt k}\right)^{-1} = 1 - e^{-2x/\sqrt k} \ , 
\end{equation}
and then, the metric is expressed as 
\begin{equation}
 ds^2 
 = - \left(1-e^{-2x/\sqrt k}\right) dt^2 + \left(1-e^{-2x/\sqrt k}\right)^{-1} dx^2 \ , 
\end{equation}
where the horizon is located at $x=0$. 
The (inverse) Hawking temperature is given by 
\begin{equation}
 \beta = 2\pi \sqrt k \ . 
\label{Hawking}
\end{equation}

The two-dimensional black hole above is reproduced 
by taking the large-dimension limit of the Schwarzschild black hole. 
The metric of the Schwarzschild black hole in the $D$-dimensional spacetime is given by 
\begin{equation}
 ds_D^2 
 = 
 - \left(1-\frac{r_h^{D-3}}{r^{D-3}}\right) d\tilde t^2 
 + \left(1-\frac{r_h^{D-3}}{r^{D-3}}\right)^{-1} dr^2 
 + r^2 d \Omega_{D-2}^2 \ , 
\end{equation}
where $d \Omega_{D-2}^2$ is the metric of the $(D-2)$-dimensional unit sphere. 
We define the radial coordinate $x$ as 
\begin{equation}
 \frac{2x}{\sqrt k}
 = (D-3)\log \frac{r}{r_h} \ . 
\end{equation}
In order to keep $x$ finite in the large-$D$ limit, 
$r$ must be infinitesimally close to $r_h$. 
It can be expanded near the horizon as 
\begin{equation}
 \frac{r}{r_h} 
 = 1 + \frac{1}{D-3} \frac{2x}{\sqrt k} + \cdots \ . 
\end{equation}
Also, the (inverse) Hawking temperature of the $D$-dimensional Schwarzschild black hole is given by 
\begin{equation}
 \beta_D = \frac{4\pi r_h}{D-3} \ , 
\end{equation}
implying that the time coordinate $\tilde t$ in the $D$-dimensional spacetime 
is related to the time in the two-dimensional black hole as 
\begin{equation}
 \tilde t 
 = \frac{\beta_D}{\beta} t \ . 
\end{equation}
Thus, the two-dimensional black hole \eqref{2dBH} describes 
the near horizon geometry of the Schwarzschild black hole in the large-dimension limit as 
\begin{equation}
 ds_D^2 
 \simeq 
 \frac{1}{k}\left(\frac{2r_h}{D-3}\right)^2 ds^2 + r_h^2 d \Omega_{D-2}^2 
 \ . 
\end{equation}


\subsection{Winding strings in two dimensions}
\label{ssec:cls-str}

Now, we review the Horowitz-Polchinski model in two-dimensional dilaton gravity \cite{Brustein:2021qkj}. 
The two-dimensional black hole geometry \eqref{2dBH} appears as 
the target space of the gauged WZW model of the $SL(2,\mathbb R)_k/U(1)$ coset, 
and the two-dimensional dilaton gravity theory \eqref{CGHS} is nothing but 
the target-space effective theory of the gauged WZW model. 
Although the action \eqref{CGHS} includes no effective field of winding strings, 
the original gauged WZW model contains them. 
Hence, it is straightforward to incorporate them into the model. 
Thus, we introduce the effective field $\chi$ of winding strings wrapping on the Euclidean time circle, 
and the action is given by 
\begin{equation}
 \mathcal I_\text{matter} 
 = 
 - \frac{1}{\pi} \int d^2 x \sqrt{-g}\,
 e^{-2\phi} \left[
  \left(\partial\chi\right)^2 
  + \frac{1}{4\pi^2} \left(\beta^2\left|g_{tt}\right| - \beta_H^2\right) \chi^2 
 \right] 
 \ , 
\label{Iwc}
\end{equation}
where $\beta$ is the inverse temperature, or equivalently the period of the Euclidean time circle, 
and $\beta_H$ is the inverse Hagedorn temperature of strings, $\beta_H^2 = 8\pi \alpha'$ \cite{Atick:1988si}. 
Hereafter, we will set $\alpha'=1$. 

By using the tortoise coordinate, the metric is expressed as 
\begin{equation}
 ds^2 = e^{2\rho} \left(- dt^2 + dr_*^2\right) \ . 
\label{metric1}
\end{equation}
We focus on static solutions and hence assume that the solution 
does not have $t$-dependence. 
It should be noted that we can no longer take $\rho = \phi$ as 
the $t$-coordinate is fixed so that it is associated to the Killing vector. 
The energy-momentum tensor of winding strings is given by 
\begin{align}
 e^{2\phi}\, T_{tt} 
 &= 
 \left(\partial_* \chi\right)^2 
 + \frac{1}{4\pi^2} \left(3 e^{4\rho} \beta^2 - e^{2\rho} \beta_H^2 \right) \chi^2 
 \ , 
\\
 e^{2\phi}\, T_{**} 
 &= 
 \left(\partial_* \chi\right)^2 
 - \frac{1}{4\pi^2} \left(e^{4\rho} \beta^2 - e^{2\rho} \beta_H^2 \right) \chi^2 
 \ , 
\\
 T_{t*} 
 &= 0 
 \ , 
\end{align}
where we assumed that the winding string field $\chi$ does not depend on $t$, 
and the index $*$ stands for $r_*$-components. 
The variation of the action with respect to the dilaton is calculated as 
\begin{equation}
 L 
 = 
 - \frac{\pi}{4\sqrt{-g}} \frac{\delta\mathcal I_\text{matter}}{\delta\phi}
 = 
 - \frac{1}{2} \left(\partial_* \chi\right)^2 
 - \frac{1}{8\pi^2} e^{2\rho-2\phi}\left(e^{4\rho} \beta^2 - e^{2\rho} \beta_H^2 \right) \chi^2 \ . 
\end{equation}
After some algebras, the equations of motion for the metric and dilaton are obtained as 
\begin{align}
 0 &= 
 2 \partial_* \phi \partial_* \rho - \partial_*^2 \rho 
 + \frac{1}{2\pi^2} e^{4\rho} \beta^2 \chi^2 
\ , 
\label{eom-tt}
\\
 0 &= 
 2 \partial_* \phi \partial_* \rho - 2 \partial_*^2 \phi + \partial_*^2 \rho 
 + 2 \left(\partial_* \chi\right)^2 
\ , 
\label{eom-rr}
\\
 0 &= 
 - 2 \partial_*^2 \phi + 4 \left(\partial_*\phi\right)^2 - \frac{4}{k} e^{2\rho}  
 - \frac{1}{2\pi^2} \left(2 e^{4\rho} \beta^2 - e^{2\rho} \beta_H^2 \right) \chi^2 . 
\label{eom-tr}
\end{align}
The equation of motion for the winding string field $\chi$ is given by 
\begin{equation}
 0 = 
 \partial_* \left(e^{-2\phi} \partial_*\chi \right) 
 - \frac{1}{4\pi^2} e^{-2\phi} \left(e^{4\rho} \beta^2 - e^{2\rho} \beta_H^2 \right) \chi \ . 
\label{eom-ch}
\end{equation}

It was found in \cite{Brustein:2021qkj} that 
solutions of the following first order differential equations, 
\begin{align}
 0 
 &= \partial_* \left(\rho-\phi\right) - \frac{\beta_H^2}{4\pi\beta} \ , 
\label{eqr}
\\
 0 
 &= 
 e^{-2\rho} \partial_* \phi 
 + \frac{\beta}{2\pi} \chi^2 + \frac{4\pi\beta}{k \beta_H^2} \ , 
\label{eqp}
\\
 0 
 &= 
 \partial_* \chi + \frac{\beta}{2\pi} e^{2\rho} \chi \ , 
\label{eqc}
\end{align}
also satisfy the equations of motion \eqref{eom-tt}--\eqref{eom-ch}. 
By substituting \eqref{eqc}, two of the equations of motion, \eqref{eom-tt} and \eqref{eom-rr} become 
\begin{align}
 0 &= \partial_*^2 \left(\rho - \phi\right) \ , 
\\
 0 &= \partial_* \left( e^{-2\rho} \partial_*\phi\right) + \frac{\beta}{\pi} \chi \partial_* \chi \ , 
\end{align}
which are nothing but the derivatives of \eqref{eqr} and \eqref{eqp}. 
Thus, \eqref{eom-tt} and \eqref{eom-rr} are satisfied if \eqref{eqr}--\eqref{eqc} are satisfied. 
By using \eqref{eqr} and \eqref{eqc}, the other equation \eqref{eom-tr} can be rewritten as 
\begin{align}
 0 
 &= 
 -2 e^{2\rho}
 \left[
 \partial_* \left( e^{-2\rho} \partial_*\phi\right) 
 + \frac{\beta}{\pi} \chi \partial_* \chi 
 \right] 
 - \frac{\beta_H^2}{\pi \beta} 
 \left[
 \partial_* \phi 
 + \frac{\beta}{2\pi} e^{2\rho} \chi^2 + \frac{4\pi\beta}{k \beta_H^2}e^{2\rho}
 \right] \ , 
\end{align}
where the first term is the derivative of \eqref{eqp} and the second term is \eqref{eqp} itself. 
By substituting \eqref{eqc}, the equation of motion \eqref{eom-ch} for $\chi$ can be rewritten as 
\begin{equation}
 0 = 
 2 \chi \partial_*\left(\rho-\phi\right) 
 + \partial_*\chi 
 + \frac{\beta}{2\pi} e^{2\rho} \chi - \frac{\beta_H^2}{2\pi \beta} \chi \ . 
\end{equation}
By using \eqref{eqr} and \eqref{eqc}, this equation is satisfied. 
Thus, solutions of \eqref{eqr}--\eqref{eqc} also satisfy 
the equations of motion \eqref{eom-tt}--\eqref{eom-ch}.


\subsection{Horowitz-Polchinski solution in two dimensions}
\label{ssec:cls-sol}

Now, we study a static solution of the winding string field in the two-dimensional gravity, 
by solving the first order differential equations \eqref{eqr}--\eqref{eqc}. 
In \cite{Brustein:2021qkj}, these differential equations are solved numerically. 
In this paper, we derive an analytic expression of the solution. 
We first solve eq.~\eqref{eqr} to obtain 
\begin{equation}
 \rho - \phi = \frac{\beta_H^2}{4\pi \beta} r_* \ , 
\label{r-sol}
\end{equation}
where the integration constant is absorbed by a constant shift of $r_*$. 
For later use, we define 
\begin{equation}
 \tilde r_* \equiv \frac{\beta_H^2}{4\pi \beta} r_* \ . 
\label{tilder}
\end{equation}
In order to solve the differential equations, 
we treat the winding string field $\chi$ as the spatial coordinate. 
The coordinate transformation is given by \eqref{eqc}, 
and the $r_*$-derivative is replaced by the $\chi$-derivative as 
\begin{equation}
 \partial_* \phi 
 = 
 \frac{d\chi}{dr_*} \frac{d\phi}{d\chi} 
 = 
 - \frac{\beta}{2\pi} e^{2\rho} \chi \frac{d\phi}{d\chi} \ , 
\end{equation}
and then, \eqref{eqp} becomes 
\begin{equation}
 \frac{d\phi}{d\chi} = \chi + \frac{8\pi^2}{k \beta_H^2\chi} \ , 
\end{equation}
which can be solved as 
\begin{equation}
 \phi = \frac{1}{2}\chi^2 + \alpha \log\chi + \phi_0 \ , 
\label{p-sol}
\end{equation}
where we have defined 
\begin{equation}
 \alpha = \frac{8\pi^2}{k \beta_H^2} \ . 
\label{alpha}
\end{equation}
Note that the Hagedorn temperature is a constant 
and given by $\beta_H^2 = 8\pi^2$ as we set $\alpha'=1$. 
Hence, the parameter $\alpha$ is always given by $\alpha = 1/k$. 

As the winding string field $\chi$ is treated as the coordinate now, 
the rescaled tortoise coordinate $\tilde r_*$ is a function of the coordinate $\chi$.  
By substituting \eqref{r-sol} and \eqref{p-sol} into \eqref{eqc}, 
we obtain the first order differential equation for $\tilde r_*(\chi)$,   
\begin{align}
 \frac{4\pi \beta}{\beta_H^2}\frac{d\tilde r_*}{d\chi} 
 &= 
 - \frac{2\pi}{\beta} e^{-2\tilde r_* - 2\phi_0} \chi^{-2 \alpha-1} e^{-\chi^2} \ . 
\end{align}
This differential equation is integrated as 
\begin{align}
 e^{2\tilde r_*} 
 &= 
 \frac{\beta_H^2}{2 \beta^2} e^{-2\phi_0} \left[\Gamma\left(-\alpha,\chi^2\right) + C_0\right] \ , 
\label{rs-sol}
\end{align}
where $\Gamma(a,x)$ is the incomplete gamma function, which is defined by 
\begin{equation}
 \Gamma(a,x) = \int_x^\infty dt\,t^{a-1} e^{-t} \ , 
\label{igamma}
\end{equation}
and $C_0$ is the integration constant. 
The conformal factor $\rho$ in the metric is already solved in \eqref{r-sol} 
and can be expressed as a function of $\chi$ as 
\begin{equation}
 e^{2\rho} 
 = \frac{\beta_H^2}{2 \beta^2} 
 \left[\Gamma\left(-\alpha,\chi^2\right) + C_0\right] \chi^{2\alpha} e^{\chi^2} \ . 
\label{rho-sol}
\end{equation}
Thus, we have obtained an analytic expression, \eqref{p-sol}, \eqref{rs-sol} and \eqref{rho-sol}, 
of the Horowitz-Polchinski solution in two-dimensions.  

In some cases, it would be convenient to use the proper spatial coordinate $r$ 
which is related to the tortoise coordinate $r_*$ as%
\footnote{%
The coordinate $r$ in this paper is $\rho$ in \cite{Brustein:2021qkj}. 
}
\begin{equation}
 dr = e^\rho dr_* \ , 
\end{equation}
and to the coordinate $\chi$ as 
\begin{align}
 dr 
 &= 
 - \frac{2\pi}{\beta} e^{-\rho} \chi^{-1} d\chi \ . 
\end{align}
The coordinate $r$ can be expressed in terms of the integral as 
\begin{align}
 r 
 &= 
 - \frac{2\sqrt 2 \pi}{\beta_H} 
 \int 
 \frac{\chi^{-\alpha-1}e^{-\frac{1}{2}\chi^2}d\chi}
  {\sqrt{\Gamma\left(-\alpha,\chi^2\right)+C_0}} \ .  
\label{r-int}
\end{align}

Now we examine the basic behavior of this solution. 
The incomplete gamma function behaves as 
\begin{align}
 \Gamma(-\alpha,\chi^2) 
 &\simeq \frac{1}{\alpha} \chi^{-2\alpha} \to \infty 
 \ , 
&
 \text{in}&\qquad
 \chi\to 0 
 \ , 
\\
 \Gamma(-\alpha,\chi^2) 
 &\simeq e^{-\chi^2}\chi^{-2\alpha-2} \to 0 
 \ , 
&
 \text{in}&\qquad
 \chi\to \infty
 \ . 
\end{align}
As the integrant of \eqref{igamma} is always positive, 
$\Gamma(-\alpha,\chi^2)$ monotonically decreases as $\chi$ increases 
and runs from the infinity to zero. 
Around $\chi = 0$, spatial coordinates $r$ and $r_*$ behave as 
\begin{align}
 r &\simeq - \log\chi \ , 
 &
 r_* &\simeq - \frac{\alpha \beta}{2\pi} \log \chi \ , 
\end{align}
where we used $\beta_H^2 = 8\pi^2$. 
Thus, the region with larger $r$, or equivalently $r_*$, 
would be identified with the outer region, where the spacetime is approximately flat. 
The metric near the outer boundary $r_*\to +\infty$ can be expanded as 
\begin{equation}
 e^{2\rho} 
 \simeq 
 \frac{4\pi^2}{\alpha\beta^2} 
 \left\{
  1 
  + \frac{4\pi^2}{\alpha\beta^2} e^{-2\phi_0}
   \left[\Gamma(-\alpha) + C_0\right] e^{-4\pi r_*/\beta} 
  + \cdots
 \right\} \ , 
\end{equation}
where we assumed that $1/\alpha = k > 1$, and then, is approximated as  
\begin{equation}
 ds^2 
 \simeq 
 \frac{4\pi^2 k}{\beta^2}\left(1 + e^{4\pi r_*/\beta}\right)^{-1} 
 \left(-dt^2 + dr_*^2\right) \ , 
\label{asympt}
\end{equation}
where the coefficient of the term $e^{4\pi r_*/\beta}$ is absorbed by a constant shift of $r_*$. 
By rescaling of $t$ and $r_*$ to absorb the overall factor, 
the expression above agrees with the black hole metric \eqref{2dBHrs} 
with the inverse Hawking temperature \eqref{Hawking}. 
Thus, the spacetime is asymptotically flat, and 
the solution approaches the black hole solution near the spatial infinity. 

The structure in the interior depends on the integration constant $C_0$. 
The winding string $\chi$ increases as $r_*$, or equivalently $r$ decreases. 
For $C_0>0$, $\chi$ diverges at finite $r_*$ as 
\eqref{rs-sol} is positive for $\Gamma(-\alpha,\chi^2)=0$. 
Since the integrant of \eqref{r-int} exponentially approaches zero in $\chi\to\infty$, 
the proper distance $r$ is also finite there. 
The metric and curvature also diverge in $\chi\to\infty$ as 
\begin{align}
 \rho &\simeq \frac{1}{2} \chi^2 
 \ , 
 & 
 R &\sim - \partial_\chi \left(e^{2\rho} \partial_\chi \rho\right) \sim - e^{\chi^2}
 \ . 
\end{align}
Thus, the solution has a singularity within a finite proper distance, 
and the Euclidean time circle becomes infinitely large, there. 

\begin{figure}[t]
\begin{center}
\includegraphics[scale=0.6]{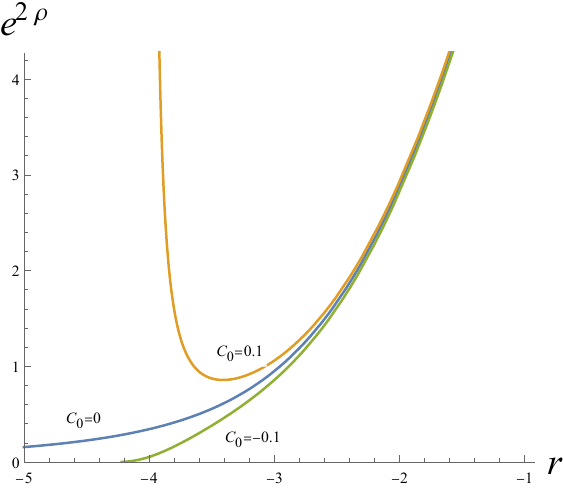}
\hspace{48pt}
\includegraphics[scale=0.6]{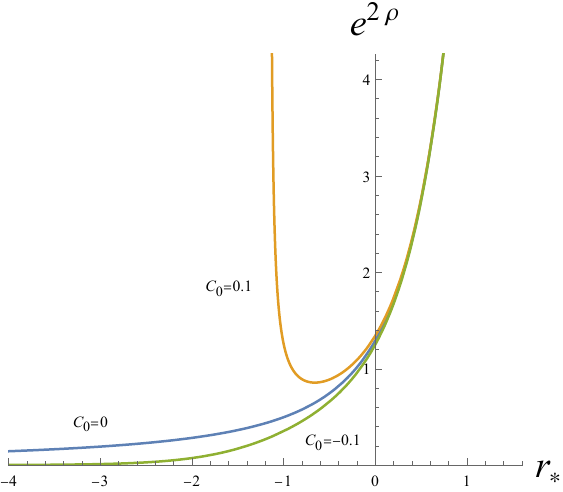}
\\[24pt]
\includegraphics[scale=0.6]{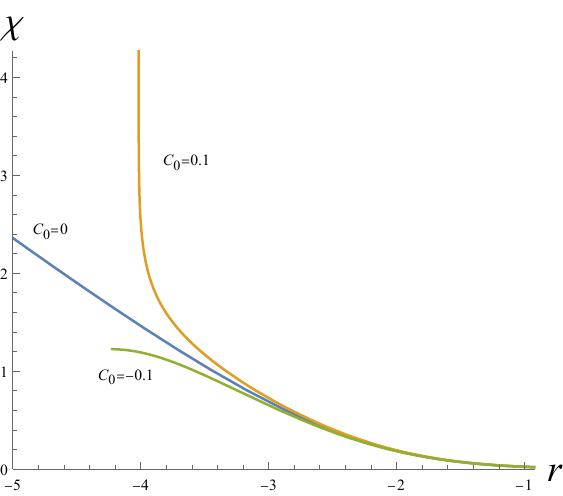}
\hspace{48pt}
\includegraphics[scale=0.6]{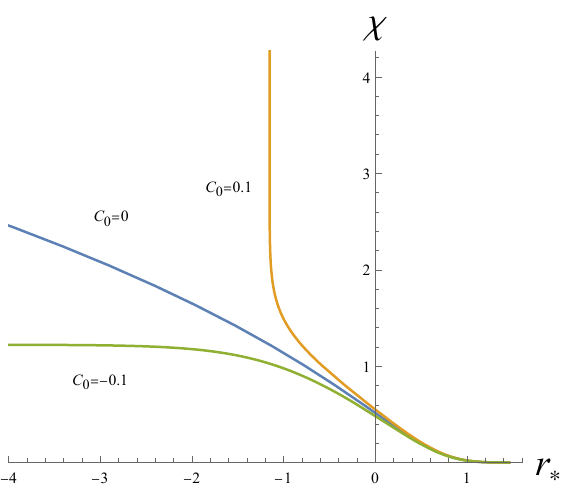}
\caption{%
Plots of classical solution $e^{2\rho}$ and $\chi$ 
in the proper coordinate $r$ and in the tortoise coordinate $r_*$. 
At a finite $r$, $e^{2\rho}$ and $\chi$ diverges for $C_0>0$, 
while $e^{2\rho}$ goes to zero for $C_0<0$. 
For $C_0=0$, both $\Omega$ and $\chi$ are finite for finite $r$. 
In the tortoise coordinate, the divergence for $C_0>0$ appears at finite $r_*$, 
while $e^{2\rho}$ approaches zero in $r_*\to-\infty$ for $C_0\leq 0$. 
}\label{fig:cl}
\end{center}
\end{figure}

For $C_0=0$, the solution has no singularity within a finite proper distance. 
The tortoise coordinate $r_*$ and the proper spatial coordinate $r$ behaves in $\chi\to\infty$ as 
\begin{align}
 r_* &\simeq - \frac{\beta}{4\pi} \chi^2 \to -\infty \ , 
 & 
 r &\simeq - \chi \to -\infty \ . 
\end{align}
Thus, the spacetime has another spatial infinity inside winding strings. 
The metric approaches zero in this limit, 
\begin{equation}
 e^{2\rho} \simeq \frac{4\pi^2}{\beta^2} \frac{1}{\chi^2} \ , 
\end{equation}
and hence, the radius of the Euclidean time circle approaches 
to zero in the inner spatial infinity. 

For $C_0<0$, the solution has the horizon. 
The tortoise coordinate reaches $r_* \to -\infty$ before $\chi$ diverges 
as \eqref{rs-sol} becomes zero for finite $\Gamma(-\alpha,\chi^2)$. 
The metric $e^{2\rho}$ is also approaching zero there, implying that it is the horizon. 
The winding string field $\chi$ takes a finite value $\chi = \chi_h$ in $r_*\to -\infty$, 
and hence, the horizon is regular. 
For the solution \eqref{rs-sol}, 
the tortoise coordinate $r_*$ is expanded around the horizon as 
\begin{align}
 e^{2\tilde r_*} 
 &= - \frac{\beta_H^2}{\beta^2} e^{-2\phi(\chi_h)} \chi_h^{-1} \left(\chi-\chi_h\right) 
 + \mathcal O \left(\left(\chi-\chi_h\right)^2\right) \ , 
\end{align}
and hence, the metric around the horizon is approximated as 
\begin{equation}
 e^{2\rho} 
 = 
 - \frac{\beta_H^2}{\beta^2} \frac{\chi-\chi_h}{\chi_h} 
 + \mathcal O \left(\left(\chi-\chi_h\right)^2\right) \ . 
\end{equation}
Near the horizon, the proper distance $r$ is related to $\chi$ as 
\begin{equation}
 \left(r - r_h\right)^2 \simeq - \frac{16\pi^2}{\beta_H^2} \frac{\chi - \chi_h}{\chi_h} \ . 
\end{equation}
The horizon $r=r_h$ is located at finite $r$, 
and the metric near the horizon is expressed in terms of $r$ as 
\begin{equation}
 ds^2 \simeq - \frac{4\pi^2}{\beta^2} \left(r-r_h\right)^2 dt^2 + dr^2 \ . 
\end{equation}
Thus, the temperature of the black hole is given by $\beta$. 
As we have seen in \eqref{asympt}, 
the time coordinate $t$ is different by a redshift factor $2\pi k/\beta$ 
from the proper time of the observer in the asymptotically flat region. 
After the rescaling of $t$, the (global) temperature is the same as 
the Hawking temperature \eqref{Hawking} in the asymptotic region. 

In summary, the solution can be classified into three cases depending on the parameter $C_0$. 
For $C_0>0$, the solution has a singularity at a finite $r$. 
For $C_0=0$, the solution is regular for finite proper distance $r$, 
and the spacetime has no horizon. 
For $C_0<0$, the spacetime has the horizon at finite $r$, 
and $\chi$ is regular at the horizon and the outside. 
Plots of the solution are shown in Fig.~\ref{fig:cl}.


\section{Self-gravitating strings in the RST model}
\label{sec:RST}


\subsection{The RST model}
\label{ssec:RST-BH}

In the black hole physics, quantum effects of matter around a black hole play an important role. 
For example, a static black hole solution is non-singular only in a specific vacuum state 
known as the Hartle-Hawking vacuum, which contains background radiations. 
Such quantum effects can be taken into account by using the RST model. 
We first review the RST model in this section. 
Assuming that the matter field has the conformal symmetry, 
the energy-momentum tensor is traceless in the classical limit 
but has the Weyl anomaly due to quantum effects. 
In curved spacetime, the trace anomaly is given by 
\begin{equation}
 \left\langle T^\mu{}_\mu \right\rangle = \frac{c}{24} R \ ,  
\end{equation}
where $c$ is the central charge. 
The expectation value of the quantum energy-momentum tensor 
can be evaluated by using the anomaly condition 
above and the conservation law $\nabla_\mu T^{\mu\nu} = 0$. 
In the double null coordinates in the conformal gauge \eqref{metric}, 
the expectation value of the energy-momentum tensor is expressed as 
\begin{align}
 \left\langle T_{\pm\pm} \right\rangle 
 &= 
 \frac{c}{12} \left[
  - \left(\partial_\pm \rho\right)^2 + \partial_\pm^2 \rho - t_\pm(x_\pm)
 \right] \ , 
\label{T++}
\\
 \left\langle T_{+-} \right\rangle 
 &= 
 - \frac{c}{12} \partial_+ \partial_- \rho 
 \ , 
\label{T+-}
\end{align}
where $t_\pm$ are the integration constants to be fixed by boundary conditions.  
The integration constants $t_\pm$ are related to the incoming and outgoing modes, respectively, 
and independent of $x_\mp$ but depends on $x_\pm$. 
The Weyl anomaly cannot be reproduced by adding any local and covariant term to the action, 
but can be obtained from a non-local or non-covariant term. 
By using the metric \eqref{metric}, the Weyl anomaly is reproduced by 
\begin{equation}
 \mathcal I_\text{anom} 
 = 
 - \frac{c}{12\pi} \int dx^2 \left(\partial_+\rho\right)\left(\partial_-\rho\right) \ . 
\end{equation}
As we have seen in the previous section, 
the two-dimensional dilaton-gravity theory consists of two fields, $\rho$ and $\phi$. 
To solve the equations of motion, it is important 
that the combination $\rho-\phi$ behaves as a free field. 
This structure comes from the fact that the classical action \eqref{CGHS} 
is invariant under the transformation 
\begin{equation}
 \delta \phi = \delta \rho = \epsilon e^{2\phi} \ . 
\label{sym}
\end{equation}
The anomaly term above, however, breaks this symmetry. 
In order to preserve the symmetry, 
an additional local and covariant term 
\begin{equation}
 \mathcal I_\text{RST} 
 = - \frac{c}{48\pi} \int dx^2 \sqrt{-g}\, \phi R 
 = - \frac{c}{12\pi} \int dx^2 \, \phi\, \partial_+ \partial_- \rho 
\end{equation}
is introduced in \cite{Russo:1992ax}. 
Then, the total action is now invariant under the modified symmetry, 
\begin{equation}
 \delta\phi = \delta\rho = \epsilon \left(e^{-2\phi} - \frac{c}{48}\right)^{-1} \ . 
\label{qsym}
\end{equation}
The equations of motion for the metric components in the conformal gauge \eqref{metric} 
are given by 
\begin{align}
 0 &= 
 \left(e^{-2\phi} - \frac{c}{48} \right)
 \left[
  4 \left(\partial_+ \rho\right) \left(\partial_+ \phi\right) - 2 \partial_+^2 \phi
 \right] 
\notag\\
 &\quad
 + T_{++} 
 + \frac{c}{12} \left[
  - \left(\partial_\pm \rho\right)^2 + \partial_\pm^2 \rho - t_\pm
 \right] \ , 
\\
 0 
 &= 
 e^{-2\phi} 
 \left[
  2 \partial_+ \partial_- \phi 
  - 4 \left(\partial_+\phi\right) \left(\partial_-\phi\right)
  - \frac{1}{k} e^{2\rho} 
 \right] 
\notag\\
 &\quad
 + T_{+-} 
 - \frac{c}{12} \partial_+ \partial_- \rho + \frac{c}{24} \partial_+ \partial_- \phi
 \ . 
\end{align}
where $T_{\mu\nu}$ are the classical part of the energy-momentum tensor, and for the dilaton $\phi$, 
\begin{align}
 0 &= 
 e^{-2\phi}\left[
  - 4 \partial_+ \partial_- \phi + 4 \left(\partial_+\phi\right) \left(\partial_-\phi\right) 
  + 2 \partial_+ \partial_- \rho + \frac{1}{k} e^{2\rho} 
 \right]
\notag\\
 &\quad 
 + e^{2\rho} L
 + \frac{c}{24} \partial_+ \partial_- \rho \ , 
\end{align}
where $L$ is given by \eqref{L} for classical matter fields. 

It is convenient to define $\Omega$ and $X$ as 
\begin{align}
 \Omega 
 &= 
 e^{-2\phi} + \frac{\kappa}{2}\phi \ , 
 & 
 X 
 &= 
 e^{-2\phi} 
 + \kappa \rho - \frac{\kappa}{2} \phi \ , 
\label{ox-def}
\end{align}
where $\kappa = {c}/{12}$. 
Then, the action is expressed in terms of $\Omega$ and $X$ as 
\begin{align}
 \mathcal I
 &= 
 \frac{1}{\pi \kappa} \int d^2 x\, 
 \left[
  - \partial_+ X \partial_- X + \partial_+ \Omega \partial_- \Omega 
  + \frac{\kappa}{k} e^{2(X-\Omega)/\kappa}
 \right] 
 + \mathcal I_\text{matter} \ . 
\label{RSTaction}
\end{align}
In the absence of the classical matter, $\mathcal I_\text{matter}=0$, 
the equations of motion are expressed as 
\begin{align}
 0 &= 
 \partial_+ \partial_- X + \frac{1}{k} e^{2\left(X-\Omega\right)/\kappa} \ , 
\label{RSTX}
\\
 0 &= 
 \partial_+ \partial_- \Omega + \frac{1}{k} e^{2\left(X-\Omega\right)/\kappa} \ , 
\label{RSTO}
\\
 0 &= 
 - \partial_\pm X \partial_\pm X + \kappa \partial_\pm^2 X 
 + \partial_\pm \Omega \partial_\pm \Omega - \kappa^2 t_\pm \ . 
\label{RSTC}
\end{align}

As in the classical case, we can take $\Omega=X$ in the RST model. 
From \eqref{RSTX} and \eqref{RSTO}, we obtain 
\begin{equation}
 \partial_+ \partial_- \left(X - \Omega\right) = 0 \ . 
\end{equation}
Since $\Omega$ is invariant under the coordinate transformation, while 
$X$ transforms, just like \eqref{transf:rho}, as 
\begin{equation}
 X \to X - \frac{\kappa}{2} \log f_+(x_+) - \frac{\kappa}{2} \log f_-(x_-) \ , 
\end{equation}
Thus, the fields $\Omega$ and $X$ can always satisfy the condition $\Omega=X$ 
by taking appropriate gauge condition, or equivalently, coordinates. 
In this gauge, the equations of motion reduce to 
\begin{align}
 \partial_+ \partial_- \Omega &= - \frac{1}{k} \ , 
 & 
 \partial_\pm^2 \Omega &= \kappa^2 t_\pm \ . 
\end{align}
The first equation is solved by 
\begin{equation}
 \Omega = - \frac{1}{k} x_+ x_- + F_+(x_+) + F_-(x_-) \ , 
\end{equation}
where $F_\pm$ are the integration constants. 
The second equation simply relates the integration constants $F_\pm$ and $t_\pm$ as 
\begin{equation}
 F_\pm''(x_\pm) = \kappa^2 t_\pm(x_\pm) \ . 
\end{equation}
Thus, the solution above with any $F_\pm$ is a vacuum solution, 
but different integration constants correspond to different vacua, 
such as Hartle-Hawking vacuum, Boulware vacuum or Unruh vacuum. 
The Hartle-Hawking vacuum corresponds to $F_\pm=\frac{1}{2}M$ with some constant $M$, 
and then, we have 
\begin{equation}
 \Omega = M - \frac{1}{k} x_+ x_- \ . 
\label{RSTBH}
\end{equation}

Although the black hole solution \eqref{RSTBH} has the same form 
as the classical solution \eqref{CGHSBH}, 
it contains non-zero quantum corrections as 
the new field $\Omega$ is related to the original dilaton $\phi$ by \eqref{ox-def}. 
In fact, the geometry has the singularity at a slightly different position 
from the classical solution where $\Omega$ is regular and non-zero. 
It is straightforward to see that the curvature has the singularity if 
\begin{equation}
 \frac{d \Omega}{d \phi} = - 2 e^{-2\phi} + \frac{\kappa}{2} = 0 \ , 
\end{equation}
and there, $\Omega$ takes the critical value 
\begin{equation}
 \Omega_c = \frac{\kappa}{4}\left(1-\log \frac{\kappa}{4}\right) \ . 
\label{oc}
\end{equation}
Since the dialton $\phi$ must be real, $\Omega$ has the lower bound at the critical value above.


\subsection{Winding strings in the RST model}
\label{ssec:RST-str}

Now, we study self-gravitating strings by taking quantum effects of matter into account. 
Self-gravitating strings are described by using winding strings wrapping around the Euclidean time circle 
as in the classical case we studied in the previous section. 
Since such winding strings at high temperatures are in the opposite limit of the worldsheet moduli space to massless strings, 
we assume that quantum effects of winding strings are suppressed, 
and that it is sufficient to take quantum effects of massless modes into calculations. 
Thus, we study quantum effects in the self-gravitating string solution 
by introducing the winding string field $\chi$ into the RST model. 
As in the RST model, we make a small modification at the quantum level 
also in the effective action of the winding string field. 
In the RST model, $\Omega$ consists only of the dilaton $\phi$ and 
can be interpreted as a quantum version of $e^{-2\phi}$. 
We assume that the effective action of the winding string field 
can also be expressed only in terms of $\Omega$, $X$ and $\chi$, 
and modify it as 
\begin{equation}
 \mathcal I_\text{matter} 
 = 
 \frac{1}{\pi} \int d^2 x\, 
 \left[
  2 \Omega \partial_+ \chi \partial_- \chi 
  - \frac{1}{8\pi^2} e^{\frac{2}{\kappa}(X-\Omega)} 
  \left(
   \Omega^{-1} e^{\frac{2}{\kappa}(X-\Omega)} \beta^2 - \beta_H^2
  \right) \chi^2 
 \right] \ , 
\label{Iw}
\end{equation}
where all correction terms are of order of quantum effects, namely of $\mathcal O(\kappa)$, 
and the action reduces to the classical action \eqref{Iwc}. 

We consider the metric in the conformal gauge \eqref{metric1} and focus on the static configurations. 
For later convenience, we define $\Psi$ as 
\begin{equation}
 \kappa \Psi = X - \Omega \ . 
\end{equation}
Then, assuming that the fields are independent of $t$, 
the action is expressed as 
\begin{align}
 \mathcal I &= 
 \frac{1}{\pi} \int dt dr_*\, 
 \left[
  \frac{\kappa}{2} \left(\partial_* \Psi \right)^2 
  + \left(\partial_* \Psi \right)\left(\partial_* \Omega \right) 
  + \frac{2}{k} e^{2\Psi}
 \right] 
\notag\\
&\quad
 - \frac{1}{\pi} \int dtdr_*\, 
 \left[
  \Omega \left(\partial_* \chi\right)^2 
  + \frac{1}{4\pi^2} e^{2\Psi} \left(\Omega^{-1} e^{2\Psi} \beta^2 - \beta_H^2\right) \chi^2 
 \right] 
 \ , 
\label{action}
\end{align}
where we used 
\begin{equation}
 d^2 x = dx_+ dx_- = 2 dt dr_* \ . 
\end{equation}
The equations of motion for $\Omega$, $\Psi$ and $\chi$ are given by 
\begin{align}
 0 
 &= 
 - \partial_*^2 \Psi - \left(\partial_* \chi\right)^2 
 + \frac{\beta^2}{4\pi^2}\Omega^{-2} e^{4\Psi} \chi^2 
 \ , 
\label{eomob}
\\
 0 
 &= 
 - \partial_*^2 \Omega - \kappa \partial_*^2 \Psi 
 + \frac{4}{k} e^{2\Psi} 
 - \frac{4\beta^2}{4\pi^2} \Omega^{-1} e^{4\Psi} \chi^2 + \frac{2\beta_H^2}{4\pi^2} e^{2\Psi}\chi^2 
 \ , 
\label{eompb}
\\
 0 
 &= 
 - \partial_*\left(\Omega \partial_* \chi \right) 
 + \frac{\beta^2}{4\pi^2} \Omega^{-1} e^{4\Psi} \chi - \frac{\beta_H^2}{4\pi^2} e^{2\Psi} \chi 
 \ . 
\label{eomcb}
\end{align}
The constraint equations which come from 
the equations of motion for the off-diagonal components of the metric become
\begin{align}
 0 &= 
 - \kappa^2 \left(\partial_*\Psi\right)^2 + \kappa \partial_*^2 \Psi + \kappa \partial_*^2 \Omega 
 - 2 \kappa \left(\partial_*\Psi\right) \left(\partial_* \Omega\right) - 4 \kappa^2 t_\pm 
\notag\\
 &\quad 
 + 2 \left(\partial_*\chi\right)^2 + \frac{\kappa \beta^2}{2\pi^2} \Omega^{-1} e^{4\Psi} \chi^2 
\ . 
\label{const}
\end{align}

In a similar fashion to the classical case, 
the equations of motion above are solved by inspecting 
the following first order differential equations, 
\begin{align}
 0 
 &= \partial_* \Psi - \frac{\beta_H^2}{4\pi\beta} \ , 
\label{eqpb}
\\
 0 
 &= 
 - \frac{1}{2} e^{-2\Psi} \partial_* \Omega 
 + \frac{\beta}{2\pi} \chi^2 + \frac{4\pi\beta}{k \beta_H^2} \ , 
\label{eqob}
\\
 0 
 &= 
 \partial_* \chi + \frac{\beta}{2\pi} \Omega^{-1} e^{2\Psi} \chi \ . 
\label{eqcb}
\end{align}
First, by differentiating \eqref{eqpb}, we obtain 
\begin{equation}
 \partial_*^2 \Psi = 0 \ . 
\label{eqp2}
\end{equation}
By substituting \eqref{eqp2}, \eqref{eomob} is satisfied by using \eqref{eqcb}. 
By substituting \eqref{eqcb} and \eqref{eqp2}, 
\eqref{eomob} becomes 
\begin{equation}
 0 = 
 - \frac{1}{2} e^{-2\Psi} \partial_*^2 \Omega 
 + \frac{2}{k}  
 + \frac{2\beta}{2\pi} \chi \partial_* \chi + \frac{\beta_H^2}{4\pi^2} \chi^2 \ .  
\end{equation}
By using \eqref{eqpb}, it can be rewritten as 
\begin{equation}
 0 = 
 - \frac{1}{2} \partial_* \left(e^{-2\Psi} \partial_* \Omega\right) 
 + \frac{\beta}{\pi} \chi \partial_* \chi 
 + \frac{\beta_H^2}{2\pi \beta} \left[
 - \frac{1}{2} e^{-2\Psi} \left(\partial_*\Omega\right) 
 + \frac{\beta}{2\pi} \chi^2 + \frac{4\pi \beta}{k \beta_H^2} \right] \ , 
\end{equation}
where the first two terms are the derivative of \eqref{eqob}, 
and the third term is nothing but \eqref{eqob} itself. 
Thus, \eqref{eompb} is satisfied by solutions of \eqref{eqpb}--\eqref{eqcb}. 
By differentiating \eqref{eqcb}, we obtain 
\begin{align}
 0 
 &= 
 \partial_*\left(\Omega \partial_* \chi\right) 
 + \frac{\beta}{2\pi} e^{2\Psi} \partial_* \chi + \frac{2\beta}{2\pi} e^{2\Psi} \chi \partial_*\Psi 
\ . 
\end{align}
Then, \eqref{eomcb} is reproduced by substituting \eqref{eqpb} and \eqref{eqcb}. 
Thus, \eqref{eomcb} is also satisfied. 
By substituting \eqref{eqpb} and \eqref{eqcb}, 
the constraint equation \eqref{const} becomes 
\begin{equation}
 0 
 = 
 - \left(\frac{\kappa \beta_H^2}{4\pi \beta}\right)^2 
 + \kappa e^{2\Psi}\partial_* \left(e^{-2\Psi} \partial_* \Omega\right) 
 - \frac{2\kappa \beta}{\pi} \Omega^{-1} e^{2\Psi} \chi \partial_* \chi - 4 \kappa^2 t_\pm \ .  
\end{equation}
By using \eqref{eqob}, we obtain the condition for 
the integration constants $t_\pm$ of the conservation law as 
\begin{equation}
 t_+ = t_- = - \left(\frac{\beta_H^2}{8\pi \beta}\right)^2 \ . 
\label{tpm}
\end{equation}


\subsection{Horowitz-Polchinski solution in the RST model}
\label{ssec:RST-sol}

Now, we solve the first order differential equations \eqref{eqpb}--\eqref{eqcb} 
to obtain a static solution of winding strings in the RST model. 
By integrating \eqref{eqpb}, we obtain 
\begin{equation}
 \Psi = \frac{\beta_H^2}{4\pi \beta} r_* = \tilde r_* \ , 
\label{ps-sol}
\end{equation}
where the rescaled tortoise coordinate $\tilde r_*$ is defined by \eqref{tilder}. 
As in the classical case, we use the winding string field $\chi$ as a spatial coordinate. 
The coordinate transformation is given by \eqref{eqcb}, 
and the $r_*$-derivative in \eqref{eqob} is replaced by the $\chi$-derivative as 
\begin{equation}
 \partial_* \Omega
 = 
 \frac{d\chi}{dr_*} \frac{d \Omega}{d\chi} 
 = 
 - \frac{\beta}{2\pi} e^{2\tilde r_*} \frac{\chi}{\Omega} \frac{d \Omega}{d\chi} \ . 
\end{equation}
Then, \eqref{eqob} becomes 
\begin{equation}
 \frac{d \log\Omega}{d\chi} = - 2\chi - \frac{16\pi^2}{k \beta_H^2\chi} \ , 
\end{equation}
which is solved as 
\begin{equation}
 \Omega = \Omega_0 \chi^{-2\alpha} e^{-\chi^2} \ , 
\label{o-sol}
\end{equation}
where $\Omega_0$ is the integration constant, and $\alpha$ is defined by \eqref{alpha}. 
As we use $\chi$ as a coordinate, the rescaled tortoise coordinate $\tilde r_*$ 
is not treated as a coordinate but is expressed as a function of $\chi$. 
By substituting \eqref{o-sol}, \eqref{eqcb} is now interpreted as a equation for $\tilde r_*$, 
\begin{align}
 \frac{4\pi \beta}{\beta_H^2} \frac{d\tilde r_*}{d\chi} 
 &= 
 - \frac{2\pi}{\beta} e^{-2\tilde r_*} \Omega_0 \chi^{-2\alpha-1} e^{-\chi^2} \ , 
\end{align}
which can be solved as 
\begin{align}
 e^{2\tilde r_*} 
 &= 
 \frac{\beta_H^2}{2\beta^2} \Omega_0 \left[\Gamma\left(-\alpha,\chi^2\right) + C_0\right] \ , 
\label{rs-q}
\end{align}
where $\Gamma(a,x)$ is the incomplete gamma function \eqref{igamma} 
and $C_0$ is the integration constant. 
Thus, we have obtained an analytic expression of the Horowitz-Polchinski solution in the RST model. 

\begin{figure}[t]
\begin{center}
\includegraphics[scale=0.6]{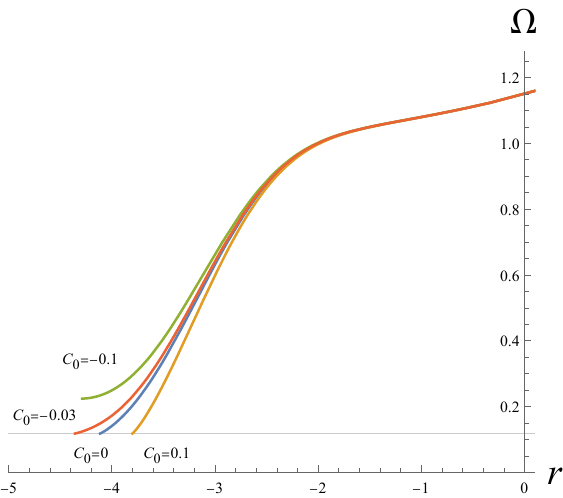}
\hspace{48pt}
\includegraphics[scale=0.6]{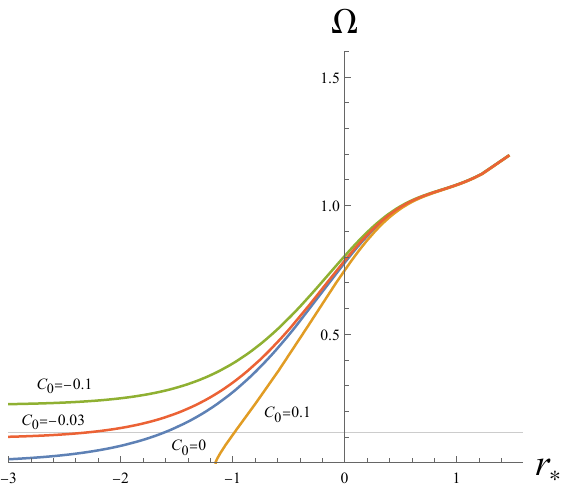}
\\[24pt]
\includegraphics[scale=0.6]{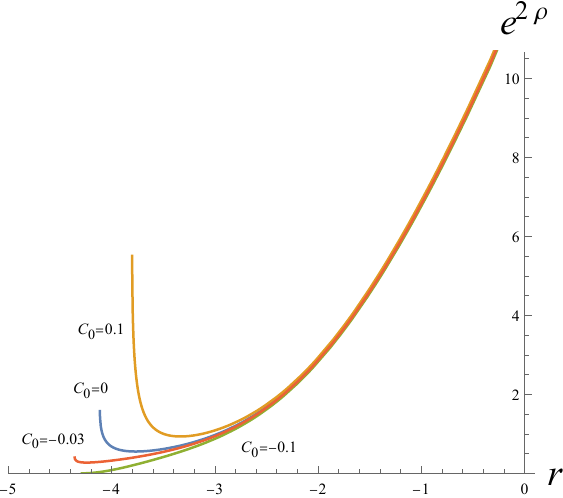}
\hspace{48pt}
\includegraphics[scale=0.6]{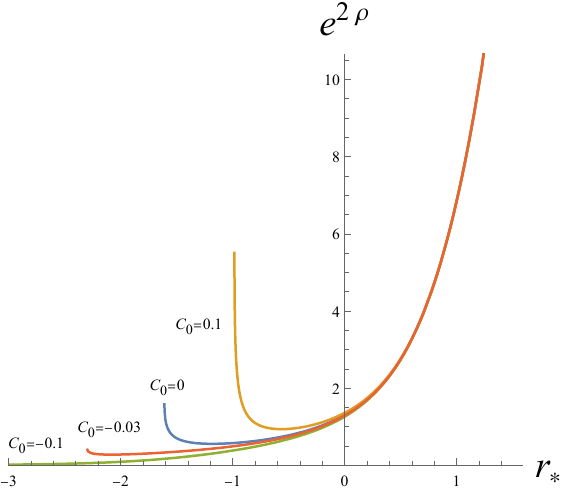}
\\[24pt]
\includegraphics[scale=0.6]{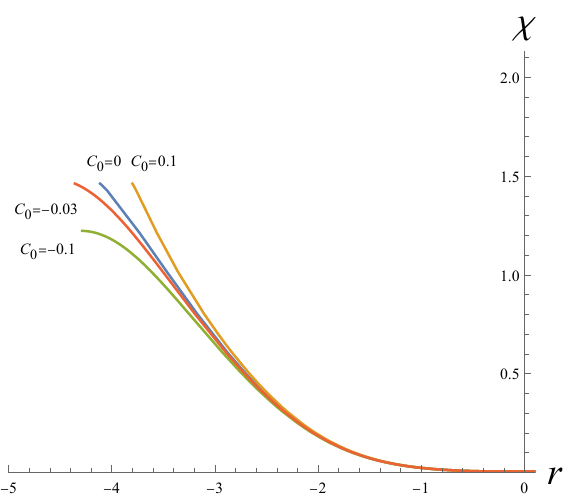}
\hspace{48pt}
\includegraphics[scale=0.6]{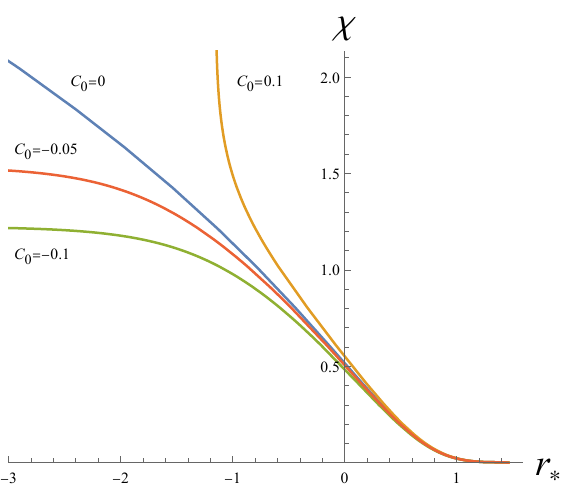}
\caption{%
Plots of solution of $\Omega$, $e^{2\rho}$ and $\chi$ in the RST model. 
The geometry has a singularity when $\Omega$ takes the critical value, $\Omega=\Omega_c$. 
In the proper coordinate $r$, solutions reach 
either the singularity $\Omega=\Omega_c$ or the horizon $e^{2\rho}=0$ at finite $r$. 
In the tortoise coordinate $r_*$, $\Omega$ goes to zero at finite $r_*$ for $C_0>0$ 
and asymptotes to its lower bound in $r_*\to-\infty$ for $C_0\leq 0$. 
The geometry has a singularity, if the lower bound is smaller than $\Omega_c$, 
or the horizon if the lower bound is larger than $\Omega_c$. 
}\label{fig:RST}
\end{center}
\end{figure}

In a similar fashion to the black hole solution, 
the solution in terms of $\Omega$, $X$ and $r_*$ have 
the same form as the solution of $\phi$, $\rho$ and $r_*$ in the classical case, 
but contains quantum corrections due to the relation between 
the RST fields $\Omega$, $X$ and original variables $\phi$, $\rho$. 
In particular, the singularity at $\Omega = \Omega_c$ plays an important role. 
In the classical case, the solution can be classified into three cases $C_0>0$, $C_0=0$ and $C_0<0$. 
The classical solution has the singularity for $C_0>0$ and the horizon for $C_0<0$ 
but neither the singularity nor horizon for $C_0=0$. 

In the RST model, we have only two cases --- 
the solution has either a singularity or a horizon. 
Since the second and third terms of \eqref{eqob} are always positive, 
$\Omega$ is a monotonically increasing function of the tortoise coordinate $r_*$. 
For $C_0\geq 0$, $\Omega$ takes the critical value \eqref{oc} at some point, 
and the geometry has the singularity there. 
The tortoise coordinate $r_*$ at the singularity is finite 
and the proper distance to the singularity is also finite. 
For $C_0<0$, it depends on $C_0$ whether $\Omega$ becomes the critical value. 
The winding string takes the maximum $\chi = \chi_h$ at $r_*\to-\infty$, 
whose value is determined by the condition 
\begin{equation}
 \Gamma\left(-\alpha,\chi^2\right) + C_0 = 0 \ . 
\label{c0-cond}
\end{equation}
As $\Omega$ is solved in \eqref{o-sol} as a monotonic function of $\chi$, 
it takes the critical value $\Omega_c$ if $\Omega(\chi_h) < \Omega_c$, 
and then, the geometry has the singularity. 
If $\Omega(\chi_h)>\Omega_c$, the spacetime has the horizon 
and $\Omega$ does not take the critical value outside the horizon. 
If $\Omega(\chi_h)=\Omega_c$, the spacetime has the horizon and is singular there. 
Provided that $\Omega_c$ is sufficiently small, \eqref{c0-cond} can be approximated as 
\begin{equation}
 \frac{1}{\alpha} \chi^{-2\alpha} + C_0 = 0 \ . 
\end{equation}
As $\Omega$ is also approximated as 
\begin{equation}
 \Omega \simeq \Omega_0 \chi^{-2\alpha} \ , 
\end{equation}
the condition for the singularity to be hidden by the horizon is approximately expressed as 
\begin{equation}
 C_0 < - k \frac{\Omega_0}{\Omega_c} \ , 
\end{equation}
where we used $\alpha = 1/k$. 

In summary, the solution has the singularity if $C_0$ is larger than some negative value 
and has the horizon if $C_0$ is smaller than that value. 
The critical point of $C_0$ is determined by 
the condition $\Omega(\chi_h) = \Omega_c$ with \eqref{c0-cond}. 
At the critical point, the geometry is singular at the horizon. 
The solutions of $\Omega$, $e^{2\rho}$ and $\chi$ are depict in Fig.~\ref{fig:RST}.


\subsection{Entropy and free energy}
\label{ssec:Entropy}

In this section, we consider the entropy and free energy of the solution. 
We first calculate the free energy $F$, 
which is related to the on-shell action as $\mathcal I = \beta F$. 
By integrating over the Euclidean time direction, 
the action has an additional overall factor of $\beta$, 
\begin{align}
 - \mathcal I &= 
 \frac{\beta}{\pi} \int d r_* \, 
 \left[
  \frac{\kappa}{2} \left(\partial_* \Psi \right)^2 
  + \left(\partial_* \Psi \right)\left(\partial_* \Omega \right) 
  + \frac{2}{k} e^{2\Psi}
 \right] 
\notag\\
&\quad
 - \frac{\beta}{\pi} \int d r_* \, 
 \left[
  \Omega \left(\partial_* \chi\right)^2 
  + \frac{1}{4\pi^2} e^{2\Psi} \left(\Omega^{-1} e^{2\Psi} \beta^2 - \beta_H^2\right) \chi^2 
 \right] 
 \ , 
\end{align}
where an additional minus factor appears due to the Wick rotation. 
Integrating the expression above by parts and substituting \eqref{eomob} and \eqref{eompb}, 
we obtain 
\begin{align}
 - \mathcal I 
 &= 
 \frac{\beta}{\pi}
 \left[
  \Omega\partial_* \Psi 
  + \frac{1}{2} \partial_* \Omega
  + \frac{\kappa}{2} \partial_* \Psi 
 \right]^{r\to\infty}_{r=r_{*\text{min}}}
 + \frac{\beta}{2\pi} \int dr_* 
 \left[
  \kappa \left(\partial_* \Psi \right)^2 
 \right] 
 \ . 
\label{osa0}
\end{align}
where $r_{*\text{min}}$ is the position of the inner boundary 
on which the geometry has either the singularity or horizon, 
and $r_{*\text{min}}=-\infty$ if it has the horizon. 
In the classical case, the on-shell action reduces to the surface term \cite{Brustein:2021qkj}, 
but in the case of the RST model, it has an additional correction term 
which cannot be written as a surface term. 
As we used only the equations of motion, 
the expression above is true for any static solutions. 

So far, we have used only the equations of motion \eqref{eomob}--\eqref{eomcb}, 
and hence, the expression \eqref{osa0} is true for any solution of the model. 
Now, we estimate the on-shell action for our solution. 
We substitute the first order differential equation \eqref{eqpb} 
to obtain 
\begin{align}
 - \mathcal I 
 &= 
 \left[
  2\Omega 
  + \frac{\beta}{2\pi} \partial_* \Omega
 \right]^\infty_{r_{*\text{min}}}
 + \frac{2\pi}{\beta} \int dr_* \,
  \kappa 
 \ , 
\label{osa}
\end{align}
Here and in the following, we use $\beta_H^2 = 8 \pi^2$. 
All terms in \eqref{osa} diverge for our solution \eqref{ps-sol}, \eqref{o-sol} and \eqref{rs-q}. 
As we will discuss later, 
the third term would be interpreted as contributions from the background radiation. 
The first two surface terms diverge at large distances $r_*\to\infty$. 
Near the infinity $r_*\to\infty$, the tortoise coordinate $r_*$ can be expanded as 
\begin{align}
 e^{2\tilde r_*} 
 &\simeq 
 \frac{4\pi^2}{\beta^2} \Omega_0 
 \left\{ \frac{1}{\alpha}\chi^{-2\alpha} + \left[\Gamma\left(-\alpha\right) + C_0\right] + \cdots \right\} \ , 
\label{rs-inf}
\end{align}
where we have assumed $1/\alpha = k > 1$, 
and hence, the winding string field $\chi$ approaches zero. 
Then, the surface terms $\Omega$ and $\partial_* \Omega$ are expanded as 
\begin{align}
 \Omega 
 &\simeq 
 \Omega_0 \chi^{-2\alpha} \ , 
\\
 \partial_* \Omega 
 &\simeq 
 \frac{4\pi}{\beta} \Omega_0 
 \left\{
  \chi^{-2\alpha} + \frac{1}{k} \left[\Gamma\left(-\alpha\right) + C_0\right]
 \right\} 
 \ . 
\end{align}
By using \eqref{rs-inf}, they can be expressed in terms of the tortoise coordinate $r_*$ as 
\begin{align}
 \Omega 
 &\simeq 
 - \frac{1}{k} \Omega_0 \left[\Gamma\left(-\alpha\right) + C_0\right] 
 + \frac{\beta^2}{4\pi^2 k} e^{2\tilde r_*} \ , 
\\
 \partial_* \Omega 
 &\simeq 
 \frac{\beta}{\pi k} e^{2\tilde r_*} \ . 
\end{align}
This asymptotic behavior near the spatial infinity is nothing but the black hole solution 
\begin{equation}
 \Omega = M + \frac{\beta^2}{4\pi^2 k} e^{2\tilde r_*} \ , 
\end{equation}
with 
\begin{equation}
 M = - \frac{1}{k} \Omega_0 \left[\Gamma\left(-\alpha\right) + C_0\right] \ . 
\label{M}
\end{equation}
The surface terms of the on-shell action \eqref{osa} in $r_*\to\infty$ behave as 
\begin{align}
 \left[-2 \Omega - \frac{\beta}{2\pi} \partial_* \Omega \right]_{r_*\to\infty} 
 &\simeq 
 - 2 M 
 - \left.\frac{\beta^2}{\pi^2 k} e^{2\tilde r_*} \right|_{r_*\to\infty} \ , 
\end{align}
and hence, diverge. 
This divergence should be regularized by subtracting the surface terms 
in the case of the flat spacetime, or equivalently for $M=0$, 
\begin{equation}
 - \mathcal I_\text{ct} 
 = \left[- 2 \Omega^\text{(flat)} - \frac{\beta}{2\pi} \partial_* \Omega^\text{(flat)} \right]_{r_*\to\infty} 
 = \left. - \frac{\beta^2}{\pi^2 k} e^{2\tilde r_*} \right|_{r_*\to\infty} \ . 
\end{equation}
By adding the counter term $\mathcal I_\text{ct}$, 
the on-shell action is given by 
\begin{align}
 \mathcal I 
 &= 
 -2 M + \left[
    2 \Omega + \frac{\beta}{2\pi} \partial_* \Omega 
   \right]_{r_*=r_{*\text{min}}} 
 - \int dr_* \frac{2\pi \kappa}{\beta} \ . 
\label{osa-reg}
\end{align}
The solution has either the singularity or horizon at the inner boundary $r_* = r_{*\text{min}}$. 
If the spacetime has the singularity, $\Omega$ takes the critical value $\Omega = \Omega_c$ on the inner boundary, 
but the expression above cannot be simplified any more. 
If the spacetime has the horizon, 
it is smooth and has no boundary at the horizon. 
There is no boundary term there, and hence 
the on-shell action is simply given by 
\begin{equation}
 \mathcal I = - 2 M - \int d r_* \frac{2\pi \kappa}{\beta} \ . 
\label{osa-rf}
\end{equation}

The on-shell action, or equivalently, the free energy 
contains only contributions from the mass of the star and the background radiation. 
Winding strings contribute the free energy only through the mass parameter. 
In fact, we can show that the on-shell action of winding strings is zero 
if the spacetime has the horizon but no singularity outside it. 
The winding string part of the action is 
\begin{equation}
 \mathcal I_w 
 = 
 - \frac{\beta}{\pi} \int d r_*\, 
 \left[
  \Omega \left(\partial_* \chi\right)^2 
  + \frac{1}{4\pi^2} e^{2\Psi} \left(\Omega^{-1} e^{2\Psi} \beta^2 - \beta_H^2\right) \chi^2 
 \right] 
 \ .  
\end{equation}
Integrating by parts and substituting the equation of motion for $\chi$ \eqref{eomcb}, 
it reduces to surface terms 
\begin{equation}
 \mathcal I_w 
 = 
 - \frac{\beta}{\pi} \left[\Omega \chi \partial_* \chi\right]^\infty_{-\infty} 
\end{equation}
By using \eqref{eqcb} and \eqref{ps-sol}, it becomes  
\begin{equation}
 \mathcal I_w 
 = 
 \frac{\beta^2}{2\pi^2} \left[ e^{2\tilde r_*}\chi^2\right]^\infty_{-\infty}  
 \ . 
\end{equation} 
This surface term vanishes at both the ends of $r_*$: At infinity $r_*\to \infty \, (\chi \to 0)$, the surface term behaves as 
\begin{equation}
 e^{2\tilde r_*} \chi^2 \simeq \chi^{-2\alpha+2} \ ,  
\end{equation}
hence vanishes and at the horizon, $\chi$ is finite but $r_*\to -\infty$ there. 
Therefore the winding string part of the on-shell action is zero if the spacetime has the horizon. 
If the spacetime has the singularity, the surface term on the singularity is finite, 
as $\chi$ and $r_*$ are finite there. 
Thus, the free energy of winding strings is always finite. 

Next, we calculate the entropy $S$, which can be calculated by using the thermodynamic relation as 
\begin{equation}
 S = \left(\beta \partial_\beta - 1\right) \mathcal I \ . 
\end{equation}
The last term of \eqref{osa} gives 
\begin{equation}
 - \left(\beta \partial_\beta - 1\right) \int dr_* \frac{2\pi \kappa}{\beta} 
 = \int d r_* \frac{4\pi \kappa}{\beta} \ , 
\end{equation}
where the volume of the space is infinite, and hence, 
we ignored the $\beta$-dependence of the volume of the space. 
This term can be interpreted as the entropy of the background energy flux of the Hartle-Hawking vacuum. 
The integration constants $t_\pm$ in the energy-momentum tensor \eqref{T++} 
can be interpreted as excitations in the flat spacetime. 
The energy density observed in the flat spacetime near the spatial infinity is 
\begin{equation}
 \varepsilon = n^\mu T_{\mu\nu} \xi^\nu \ ,  
\end{equation}
where $n^\mu$ is the unit normal vector on a timeslice and $\xi = \partial_t$. 
The energy density of the background flux for our solution \eqref{tpm} is estimated as 
\begin{equation}
 \varepsilon = \frac{1}{\pi} e^{-\rho} \left(T_{++} + T_{--}\right) = \frac{2\pi\kappa}{\beta^2} e^{-\rho} \ , 
\end{equation}
where we used $\beta_H^2 = 8\pi^2$. 
Note that our definition \eqref{EMtensor} of the energy-momentum tensor 
contains an additional factor of $\pi$. 
The entropy density is given by 
\begin{equation}
 s = \beta\left(\varepsilon + P\right) = 2 \beta \varepsilon = \frac{4 \pi\kappa}{\beta} e^{-\rho} \ , 
\end{equation}
where $P = \varepsilon$ is the pressure of the background energy flux. 
Thus, the total entropy of the background flux is expressed as 
\begin{equation}
 S_\text{flux} = \int dr_* e^{\rho} s = \int dr_* \frac{4\pi\kappa}{\beta} \ . 
\end{equation}

Now, we consider the surface terms in \eqref{osa-reg}. 
If the spacetime has the singularity, $\Omega$ takes 
the critical value $\Omega=\Omega_c$ on the inner boundary, 
and hence, is independent of $\beta$. 
Since the solution \eqref{o-sol} of $\Omega$ is 
a function only of $\chi$ (and the intengration constant $\Omega_0$), 
the winding string $\chi$ on the inner boundary is also independent of $\beta$. 
The $\beta$-dependence of $\partial_* \Omega$ can be seen 
from \eqref{eqob} with \eqref{ps-sol} and \eqref{rs-q} as 
\begin{equation}
 \partial_* \Omega \propto \frac{1}{\beta} \ . 
\end{equation}
Thus, all surface terms in \eqref{osa-reg} has no $\beta$-dependence. 
The entropy is calculated by using the thermodynamic relation as 
\begin{align}
 S 
 &= 
 2M - \left[
    2 \Omega + \frac{\beta}{2\pi} \partial_* \Omega 
   \right]_{r_*=r_{*\text{min}}} 
 + \int dr_* \frac{4\pi \kappa}{\beta} \ . 
\end{align}

If the spacetime has the horizon, the on-shell action is given by \eqref{osa-rf}. 
Since $M$ is independent of the temperature, the entropy is calculated as 
\begin{equation}
 S = 2 M + \int d r_* \frac{4\pi \kappa}{\beta} \ . 
\end{equation}
It should be noted that $M$ contains effects of winding strings 
and larger than the entropy of the black hole. 
In the case of the black hole solution, the entropy is simply given 
in terms of $\Omega$ at the horizon, 
\begin{equation}
 S_\text{BH} = 2 \Omega_h = 2M \ . 
\end{equation}
In the case of our solution, $\Omega$ 
is expressed in terms of $\chi$ as \eqref{o-sol}, 
while $M$ is given by \eqref{M}. 
By using the following formula of the gamma functions, 
\begin{align}
 \Gamma(a+1) &= a \Gamma(a) \ , 
& 
 \Gamma(a+1,x) &= a \Gamma(a,x) + x^a e^{-x} \ , 
\end{align}
it is straightforward to see that $M>\Omega_h$ because 
\begin{align}
 M 
 &= \Omega_0 \Gamma\left(1-\alpha\right) - \alpha \Omega_0 C_0 
\notag\\
 &> \Omega_0 \Gamma\left(1-\alpha,\chi_h^2\right) - \alpha \Omega_0 C_0
\notag\\
 &= 
 -\alpha \Omega_0 \Gamma\left(-\alpha,\chi_h^2\right) 
 + \Omega_0 \chi_h^{-2\alpha} e^{-\chi_h^2} 
 - \alpha \Omega_0 C_0 
\notag\\
 &= 
 \Omega_h \ , 
\end{align}
where we used \eqref{c0-cond}. 
Since $2\Omega_h$ should be interpreted as the entropy of the black hole, 
the entropy of $2M$ would contain effects of winding strings.


\section{Conclusion and discussions}
\label{sec:Conclusion}

In this paper, we studied a solution of self-gravitating strings 
and quantum effects in two-dimensional dilaton gravity. 
It was proposed in \cite{Susskind:1993ws} that a black hole transits into a self-gravitating string 
if the string coupling is taken to be very small. 
Black hole solutions in two-dimensional dilaton gravity can be interpreted as 
the near horizon geometry of the higher dimensional black hole spacetime in the large-dimension limit. 
Thus, the corresponding self-gravitating string solution in two dimensions would describe 
the geometry near the surface of a star made of self-gravitating strings in the large-dimension limit. 
A solution of self-gravitating strings in higher dimensions 
is proposed by Horowitz and Polchinski \cite{Horowitz:1997jc}. 
In this paper, we studied its counterpart in the two-dimensional dilaton gravity including quantum effects. 

The solution of self-gravitating strings in two-dimensional classical gravity 
was numerically studied in \cite{Brustein:2021qkj}. 
In this paper, we derived an analytic expression of the solution 
and investigated it in more detail. 
The solution can be classified into three cases depending on an integration constant: 
the solution with a singularity, the solution with a horizon 
and the solution without a horizon or singularity. 
In any case, the winding string field is suppressed near the spatial infinity 
in one side of the one-dimensional space 
and becomes large on the other side. 
The winding string field diverges at the inner boundary if the horizon is absent, 
but is finite if the spacetime has a horizon. 
The solution without a horizon or singularity 
would correspond to the spacetime near the surface of 
self-gravitating strings in the large-dimension limit. 
The solution with horizon would describe 
the near horizon geometry of higher-dimensional black holes 
with ``hair" of winding strings. 
The solution with singularity would not have 
higher-dimensional counterpart as it is singular. 

Although our solution is expected to correspond to 
the Horowitz-Polchinski solution in the large-dimension limit, 
it is known that there is no solution of self-gravitating strings in spacetime with dimensions higher than seven. 
Since the Horiwitz-Polchinski solution in higher dimensions 
is usually studied in the weak gravity limit, 
there would be a room for higher-dimenional Horowitz-Polchinski solution 
by taking non-linear effects of gravity into account. 
In our calculation, we did not resort to the weak gravity limit, 
and the equations of motion are solved exactly including the non-linear effects. 
It should be noted that there might be another possibility that 
the singularity cannot be seen because we focused only near the surface of the star. 
When we consider bound states in a spherically symmetric potential, 
solutions in general have a singularity at the center, 
and the condition that the singularity disappears 
restricts solutions to a discrete spectrum. 
The absence of the Horowitz-Polchinski solution with dimension larger than seven 
would imply that the singularity cannot be resolved, 
but solutions with singularity at the center would exist. 
By focusing on the surface of the bound state, 
we cannot see if there is a singularity at the center. 
Thus, we could find the Horowitz-Polchinski solutions in the large-dimension limit. 

In order to see quantum effects, we studied the Horowitz-Polchinski solution in the RST model. 
The RST model allows us to solve the equations of motion exactly including quantum effects. 
In this case, the solution can be classified into two cases, 
the solution with and without the horizon. 
The RST model always has the singularity at $\Omega = \Omega_c$. 
For the static black hole solution, this singularity is hidden by the horizon. 
Our solution with the horizon describes the black hole with a hair of winding strings, 
and the solution without the horizon corresponds to the Horowitz-Polchinski solution. 
Although the singularity appears in the RST model, 
this does not imply that the singularity in the large-dimension limit 
would be an artifact of the modification terms in the RST model. 
Our solution contains the background energy flux as a quantum effect even without the horizon. 
This energy flux behaves as the thermal radiation at the same temperature as that of winding strings.

\section*{Acknowledgments}

This work was supported in part by MEXT KAKENHI Grant-in-Aid for Transformative Research Areas
A Extreme Universe No.~JP21H05182, JP21H05186.

\end{document}